\documentclass[conference]{IEEEtran}
\IEEEoverridecommandlockouts    

\usepackage{multirow}
\usepackage{lipsum}
\usepackage{amsmath}
\usepackage{amssymb}
\usepackage[ruled,vlined]{algorithm2e}
\usepackage{graphicx}
\graphicspath{{./Figures/}}

\usepackage{tikz}
\usepackage{tikzpagenodes}
\usetikzlibrary{calc}
\usepackage{cite} 
\usepackage{xcolor} 
\usepackage{subcaption} 
\usepackage{textcomp} 
\usepackage{cuted} 
\usepackage{booktabs} 
\usepackage{multirow} 
\usepackage{svg} 

\usepackage[binary-units]{siunitx}
\makeatletter
\let\NAT@parse\undefined
\makeatother

\usepackage{xurl} 
\usepackage[hidelinks]{hyperref}

\newcommand{\fig}[1]{Fig.~\ref{#1}}

\newcommand{\tab}[1]{Table~\ref{#1}}

\newcommand{\sect}[1]{Section~\ref{#1}}

\usepackage{pict2e}
\newcommand{\colorcircle}[1]{
  {\unitlength=1ex
   \begin{picture}(1.5,1)
     \linethickness{0.1ex}
     \put(0.7,0.7){\color{#1}\circle*{1.6}}
     \put(0.7,0.7){\color{black}\circle{1.6}}
   \end{picture}}%
}

\definecolor{rwthblue}{HTML}{0071BD}
\definecolor{rwthblack}{RGB}{0,0,0}
\definecolor{rwthgrey}{RGB}{100,101,103}
\definecolor{rwthpetrol}{HTML}{006668}
\definecolor{rwthturquoise}{HTML}{00A9AC}
\definecolor{rwthgreen}{HTML}{4FB748}
\definecolor{rwthmaygreen}{HTML}{B3D334}
\definecolor{rwthyellow}{HTML}{FFF203}
\definecolor{rwthorange}{HTML}{FBA720}
\definecolor{rwthred}{HTML}{D22129}
\definecolor{rwthmagenta}{HTML}{ED0C72}
\definecolor{rwthbordeaux}{RGB}{161,16,53}
\definecolor{rwthviolet}{RGB}{97,33,88}
\definecolor{rwthpurple}{RGB}{122,111,172}

\newcommand\copyrighttext{%
    \footnotesize \copyright{ }2026 IEEE. Personal use of this material is permitted. Permission from IEEE must be obtained for all other uses, in any current or future media, including reprinting/republishing this material for advertising or promotional purposes, creating new collective works, for resale or redistribution to servers or lists, or reuse of any copyrighted component of this work in other works.}
\newcommand\copyrightnotice{%
    \begin{tikzpicture}[remember picture,overlay]
    \node[anchor=south,yshift=15pt,xshift=0pt] at (current page.south) {\parbox{\dimexpr\textwidth-\fboxsep-\fboxrule\relax}{\copyrighttext}};
    \end{tikzpicture}%
}

\title{\LARGE \bf \textit{karl.}\ -- A Research Vehicle for Automated and Connected Driving}

\author{
	\parbox{\textwidth}{%
		\centering
		Jean-Pierre Busch$^{*}$, Lukas Ostendorf$^{*}$, Guido Linden$^{*}$, Lennart Reiher$^{*}$,\\ and Till Beemelmanns$^{\dagger}$, Bastian Lampe$^{\dagger}$, Timo Woopen$^{\dagger}$, Lutz Eckstein$^{\dagger}$%
	}%
	\thanks{$^{*}$These main authors contributed equally to this work.}%
	\thanks{$^{\dagger}$Till Beemelmanns was responsible for sensor calibration. Bastian Lampe is group lead \textit{Automation}, Timo Woopen is department head \textit{Vehicle Intelligence \& Automated Driving}, and Lutz Eckstein is head of the institute.}%
    \thanks{All authors are with the Institute for Automotive Engineering (ika), RWTH Aachen University, Germany. {\tt\footnotesize \href{mailto:jean-pierre.busch@ika.rwth-aachen.de}{\{first.last\}@ika.rwth-aachen.de}}}%
}

\begin{document}
	
\maketitle
\bstctlcite{IEEEexample:BSTcontrol}
\thispagestyle{empty}
\pagestyle{empty}

\copyrightnotice
	

\begin{strip}
  \centering
  \vspace*{-2.3cm}
  \begin{tikzpicture}
    \node[inner sep=0] (img) {\includegraphics[width=\textwidth]{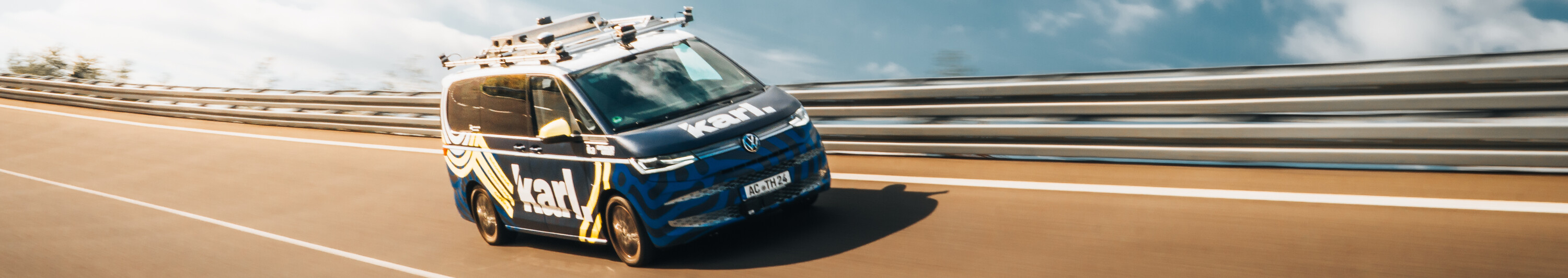}};
  \end{tikzpicture}
  \label{fig:karl}
  \vspace*{-0.8cm}
\end{strip}

\begin{abstract}
As highly automated driving is transitioning from single-vehicle closed-access testing to commercial deployments of public ride-hailing in selected areas (e.g., Waymo), automated driving and connected cooperative intelligent transport systems~(C-ITS) remain active fields of research. Even though simulation is omnipresent in the development and validation life cycle of automated and connected driving technology, the complex nature of public road traffic and software that masters it still requires real-world integration and testing with actual vehicles. Dedicated vehicles for research and development allow testing and validation of software and hardware components under real-world conditions early on. They also enable collecting and publishing real-world datasets that let others conduct research without vehicle access, and support early demonstration of futuristic use cases. In this paper, we present \textit{karl.}, our new research vehicle for automated and connected driving. Apart from major corporations, few institutions worldwide have access to their own L4-capable research vehicles, restricting their ability to carry out independent research. This paper aims to help bridge that gap by sharing the reasoning, design choices, and technical details that went into making \textit{karl.}\ a flexible and powerful platform for research, engineering, and validation in the context of automated and connected driving. More impressions of \textit{karl.}\ are available at \url{https://karl.ac}.
\end{abstract}

\section{Introduction}

Over the course of the last few years, highly automated driving as defined by SAE~Level~4~(L4)~\cite{SAEJ3016} has entered the phase of commercial operations. Today, several companies are offering publicly available ride-hailing services in selected areas. In the US, \textit{Waymo} is completing more than \num{250000}~paid autonomous rides per week in five cities~\cite{waymo}. \textit{Zoox} has launched public robotaxi services between venues in Las Vegas in 2025~\cite{zoox}. Meanwhile in China, \textit{Baidu's Apollo~Go} has recently crossed the same number of weekly autonomous rides as \textit{Waymo}~\cite{baidu}.

Nevertheless, automated or autonomous driving is still a hot topic in research with \num{10000} scientific publications per year\footnote{More than \num{10000} works per year (2022-2024) containing the term \mbox{\textit{"Autonomous Driving"}} in title or abstract, according to \href{https://openalex.org/works?page=1&filter=title_and_abstract.search:autonomous+driving&include_xpac=true}{\textit{OpenAlex}}.}. Much of the progress in the field is driven by comprehensive simulation suites, e.g., the \textit{CARLA} simulator~\cite{Dosovitskiy17,carlos}. Simulation is playing a large role in the entire development and homologation process of automated driving systems~(ADS)~\cite{Abdellatif2019-lt}. Despite substantial advancements in simulation fidelity, the so-called reality gap remains~\cite{multimodal}, mandating to also integrate, test, and validate ADS components under real-world conditions.

The availability of (partly) automated vehicles in academia has played a large role in the field's research, e.g., through the publication of large-scale datasets. However, not many institutions worldwide have access to their own L4-enabled research vehicles, severely limiting their testing and validation capabilities. One reason is that building such a research platform is both difficult and expensive, particularly because there are few role models.

In this paper, we present our new research vehicle for automated and connected driving, nicknamed \textit{karl.}\footnote{\textit{karl.} is named after \textit{\href{https://en.wikipedia.org/wiki/Charlemagne}{Charlemagne}}, emperor of the Carolingian Empire and closely related to the city of Aachen.}. We describe the key reasoning, design considerations, and technical details behind the vehicle. As a comprehensive research platform, \textit{karl.}\ is designed to serve as a role model for other automated vehicle platforms and to support progress in the field.

\section{Related Work}

This section gives an overview of L4-enabled ADS that are currently in use, with a focus on solutions and accompanying publications by other academic institutions.

\textit{Waymo} and \textit{Zoox} currently show opposing approaches with respect to the overall vehicle concept: while \textit{Waymo} is equipping series production vehicles with sensors and other necessary hardware, \textit{Zoox} has developed its vehicle from scratch, tailor-made for L4 ride-hailing with no steering wheel, and facing seats. For testing purposes, however, \textit{Zoox} is also using conventional, retrofitted vehicles, and vice-versa, \textit{Waymo} is considering purpose-built shuttles.

\textit{Zoox}-like L4 shuttles have also been built in academia: the UNICAR\textit{agil} project has produced four different vehicle prototypes (e.g., \textit{autoTAXI}) based on a common architecture, including dedicated sensor modules at all four vehicle corners~\cite{Woopen:749158}. While such non-conforming futuristic vehicle concepts offer opportunities to research not only ADS features, but also aspects like HMI or user acceptance, most automated research vehicles still focus on the actual automated driving capabilities.

The \textit{OPA\textsuperscript{3}L} project vehicle is maintained by the \textit{University of Bremen}. The customized \textit{Volkswagen Passat GTE} is "designed for fully autonomous urban driving".~\cite{opal}.

\textit{EDGAR} is \textit{TU Munich's} current research vehicle. It is similar to \textit{karl.}\ in that it is also based on a \textit{Volkswagen T7 Multivan}. The vehicle is meant to act as a comprehensive "autonomous driving research platform", including test drives in public road traffic.~\cite{karle2024edgarautonomousdrivingresearch}.

\textit{CoCar NextGen} is another multi-purpose research platform vehicle, developed by German research institution \textit{FZI} and based on an \textit{Audi A6 Avant}.~\cite{cocar}

Naming two examples outside of Germany, the \textit{University of Luxemburg} has recently built \textit{RoboCar} and the \textit{Latvian Institute of Electronics and Computer Science} has developed the \textit{White and Blue Kias}. These vehicles are based on a \textit{KIA Soul EV}. In comparison to the previously described vehicles, the hardware and sensor stacks are rather limited.~\cite{robocar, whitekia}

This brief overview only presents research vehicles with accompanying scientific publications. Looking at the extended research landscape in Germany, we notice that even more projects base off from a \textit{Volkswagen T7 Multivan}, e.g., \textit{TU Ilmenau's P:Mover}~\cite{tuilmenau} or the one by \textit{TOPAS}~\cite{topas}.

\section{Vehicle Setup of \textit{karl.}}
\label{sec:vehicle-setup}

The following section focuses on the overall hardware setup of \textit{karl.}. It details all integrated hardware components and physical modifications that enable the vehicle's automated driving features.

\subsection{Base Vehicle}
\label{ssec:base-vehicle}

\textit{karl.}\ is based on a \textit{Volkswagen T7 Multivan 1.4 TSI e-Hybrid long wheelbase}. Key criteria for selecting this model were sufficient interior space, a hybrid powertrain, and strong integration potential for automated driving research. Especially the interior's flexibility and available space are important, as they enable integrating the required compute and sensor hardware while still providing a comfortable developer workspace.
The plug-in hybrid drivetrain enables locally emission-free operation in electric mode, while at the same time, fulfilling the range and endurance requirements for extended operation. Additionally, the vehicle is equipped with driver assistance features, including active parking assist and adaptive cruise control~(ACC), which are required for the actuation of the vehicle~(\sect{sec:dbw-system}). These factors make the \textit{Volkswagen T7 Multivan} a suitable base for an automated research vehicle.

\subsection{Physical Modifications}
\label{ssec:physical-modifications}

There are only a few direct physical modifications to the vehicle: installation of power and data cables below the interior trim and feed-through to the roof; tapping the series CAN for the actuator control (\sect{sec:dbw-system}); and wrapping the car exterior (see title banner).
The main part of the modification involves installing new hardware and is described in detail in the following subsections.

\subsubsection{Cabin Rack}
\label{ssec:cabin-racl}

\begin{figure}[!t]
    \centering
    \includegraphics[width=1.0\linewidth]{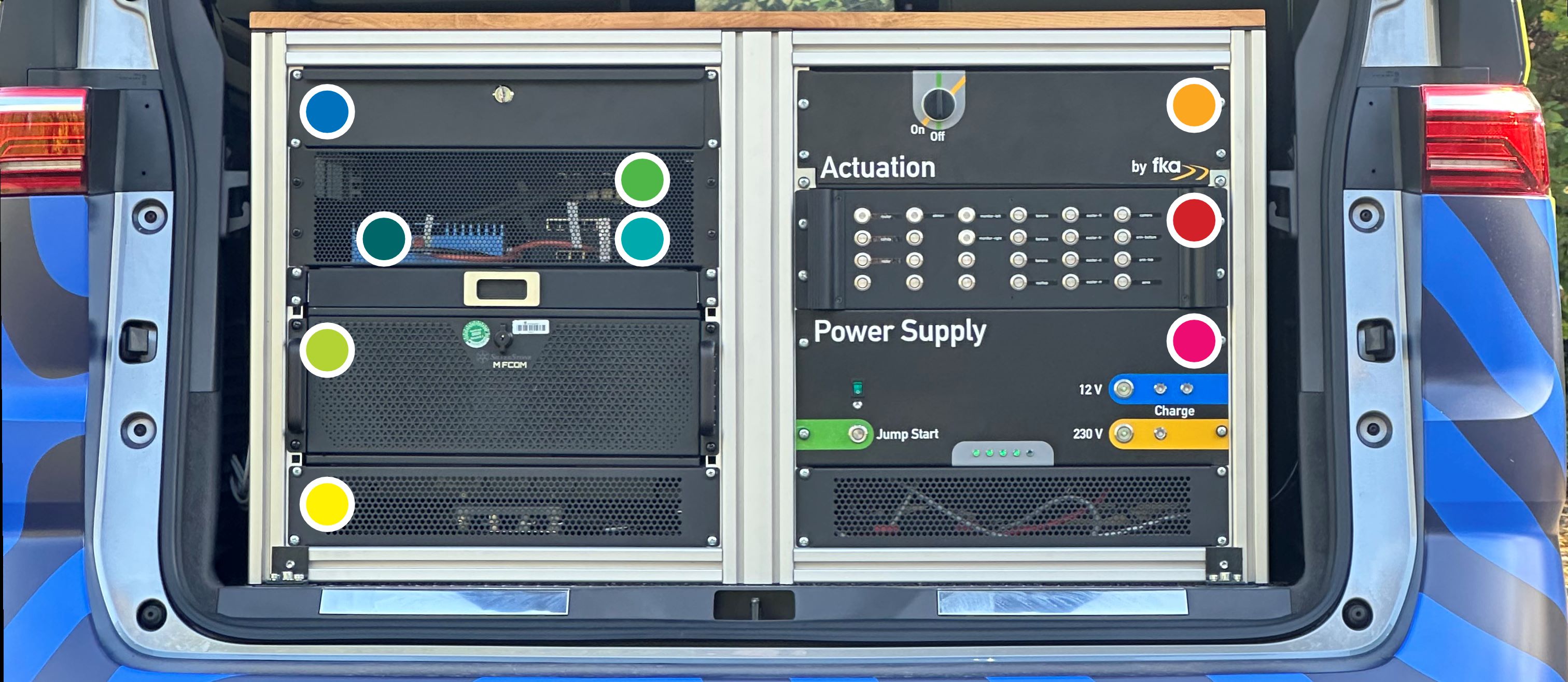}
    \caption{Cabin rack carrying \protect\colorcircle{rwthblue}~drawer, \protect\colorcircle{rwthpetrol}~V2X unit, \protect\colorcircle{rwthturquoise}~5G router, \protect\colorcircle{rwthgreen}~AE switch, \protect\colorcircle{rwthmaygreen}~HPC, \protect\colorcircle{rwthyellow}~core switch, \protect\colorcircle{rwthorange}~vehicle interface, \protect\colorcircle{rwthred}~power distribution panel, and \protect\colorcircle{rwthmagenta}~power supply}
    \label{fig:cabin-rack}
\end{figure}

Most of the hardware installed in the interior is located in a compact cabin rack in the rear compartment.
The rack is constructed from \SI{40}{mm} aluminum profiles and is divided into two separate 19" 12U columns. This provides a highly modular and expandable basis for integrating hardware components.
The rack contains the general power supply (\sect{ssec:power-supply}), a high-performance computer (HPC), two Ethernet switches (\sect{sec:compute-networking}), and external communication modems (\sect{sec:communication}), thus serving as a central data and power distribution point. In addition, further components related to sensing and actuation are integrated into the rack.
The cabin rack is shown in \fig{fig:cabin-rack}.

\subsubsection{Sensor Rack}
\label{sec:sensor-rack}

\begin{figure}[!b]
    \centering
    \includegraphics[width=1.0\linewidth]{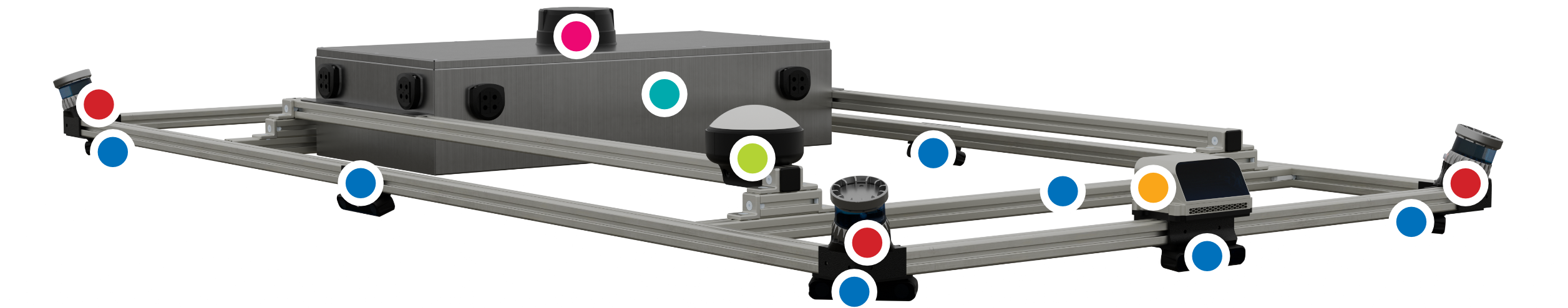}
    \caption{Sensor rack carrying \protect\colorcircle{rwthblue}~stereo cameras, \protect\colorcircle{rwthred}~rotating lidars, \protect\colorcircle{rwthorange}~FMCW lidar, \protect\colorcircle{rwthturquoise}~rooftop box, \protect\colorcircle{rwthmagenta}~5G antenna, and \protect\colorcircle{rwthmaygreen}~GNSS antennas}
    \label{fig:sensor-rack}
\end{figure}

For mounting environment sensors, we have installed a sensor rack on the roof~(\fig{fig:sensor-rack}). Compared to sensor integration in the vehicle chassis, this has the advantages of being less invasive and keeping the system modular, adaptable, and extensible.
Like the cabin rack, the sensor rack is constructed from \SI{40}{mm} aluminum profiles. It is designed to not exceed the vehicle geometry and offer flexible mounting options for an optimal field-of-view of individual sensors.

In addition to the sensors, a rooftop box is integrated into the rack. It supplies all sensors with power and consolidates their data streams at an Ethernet switch. The rooftop box also houses computing hardware that can preprocess selected sensor data~(\sect{sec:embedded-ai-computers}). The box is waterproof, actively cooled, and connected to the cabin rack via a single fiber-optic cable for transferring data, and a copper cable for power.

\subsubsection{Developer Workplace \& HMI}
\label{ssec:dev-hmi}

\begin{figure}[t]
    \centering
    \includegraphics[width=1.0\linewidth]{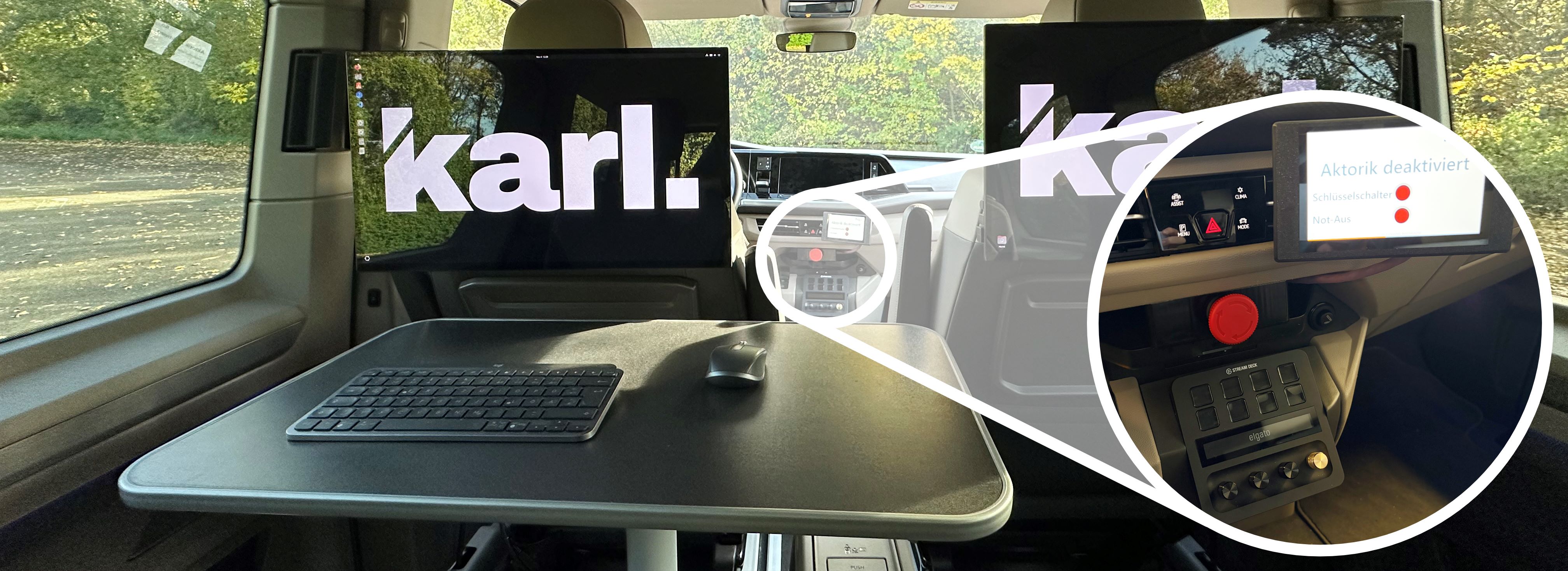}
    \caption{Developer workplace and front row HMI devices}
    \label{fig:developer-workspace}
\end{figure}

An important aspect of building a research vehicle is to create a suitable operator environment.
Two 21" touch monitors are mounted to the front seat headrests in the passenger compartment. Combined with a table mounted in the vehicle's built-in rail system, they create a comfortable developer workspace~(\fig{fig:developer-workspace}). The monitors are connected to the HPC in the cabin rack and are used for development as well as for visualizations during demonstrations.

In the front of the vehicle, two HMI components are installed: a small monitor that permanently displays the actuator system status; and an \textit{Elgato StreamDeck+} that allows the operator to interact with the software modules of our ADS~(\sect{sec:ad-stack}), e.g., for selecting the next destination or monitoring system status.

\subsection{Power Supply}
\label{ssec:power-supply}

All additionally installed hardware is powered by an isolated auxiliary power system built around a \textit{Clayton LPS~II~3000} with a \textit{CPG4} \SI{280}{Ah} expansion battery, providing a total usable capacity of approximately \SI{5}{kWh}. The power system can be charged either from the vehicle's low-voltage system (via the OEM DC/DC converter) or externally through a \SI{230}{V}~AC~inlet. It delivers up to \SI{3000}{W} continuous AC power and up to \SI{2160}{W} continuous DC power.

The system's AC output supplies the HPC (\sect{ssec:hpc}) and the core Ethernet switch (\sect{sec:networking-switches}) in the cabin rack. All remaining devices are powered from the system's DC rail via a dedicated distribution unit built around a \textit{Weidmüller UC20-WL2000-AC}. This unit provides 24 individually fused outputs that can be switched either manually or via software, allowing selective activation of subsystems during development.

One of the distribution unit's outputs feeds a high-power DC line to the rooftop box. There, a second distribution stage -- controlled by a \textit{Weidmüller UR20-FBC-MOD-TCP-ECO} via Modbus TCP from the \textit{UC20} module -- powers the rooftop sensors and compute. This stage incorporates a DC–DC step-up converter for components requiring a \SI{24}{V} rail, with all outputs again being individually fused and switchable.

Together, the two distribution layers provide high flexibility and straightforward expandability for the sensor configuration.

\subsection{Environment Sensors}
\label{ssec:environment-sensors}

The primary sensor suite for perception of the vehicle's environment is composed of cameras, lidars, and radars. The cameras provide high-resolution visual detail, the corner lidars supply precise 3D geometry, and the front-facing FMCW lidar additionally provides direct velocity information. The radars complement these modalities with further distance and velocity measurements.

\begin{figure*}[t]
  \centering
  \begin{subfigure}[b]{0.665\textwidth}
    \includegraphics[width=\textwidth]{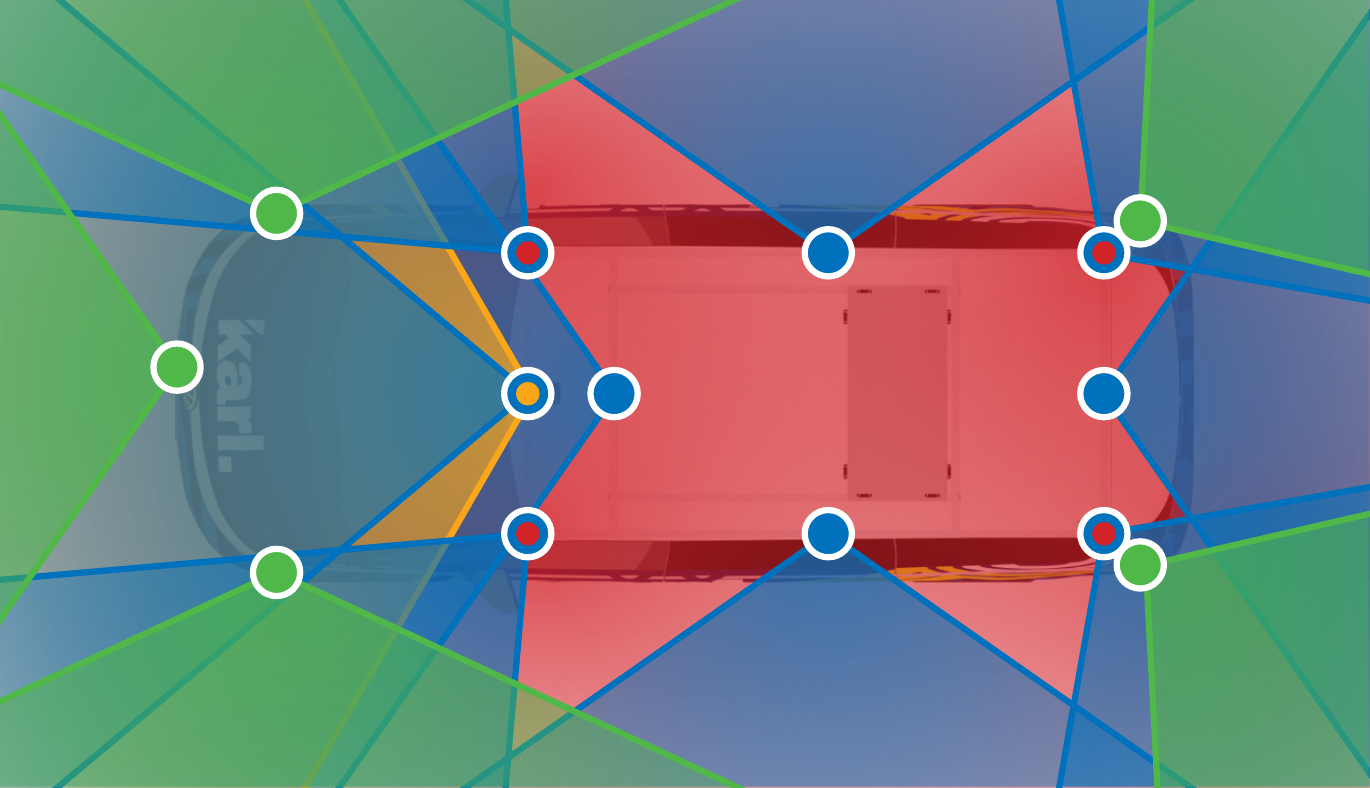}
    \caption{Top-down view including all cameras, lidars, and radars}
  \end{subfigure}
  \hfill
  \begin{subfigure}[b]{0.315\textwidth}
    \centering
    \includegraphics[width=\textwidth]{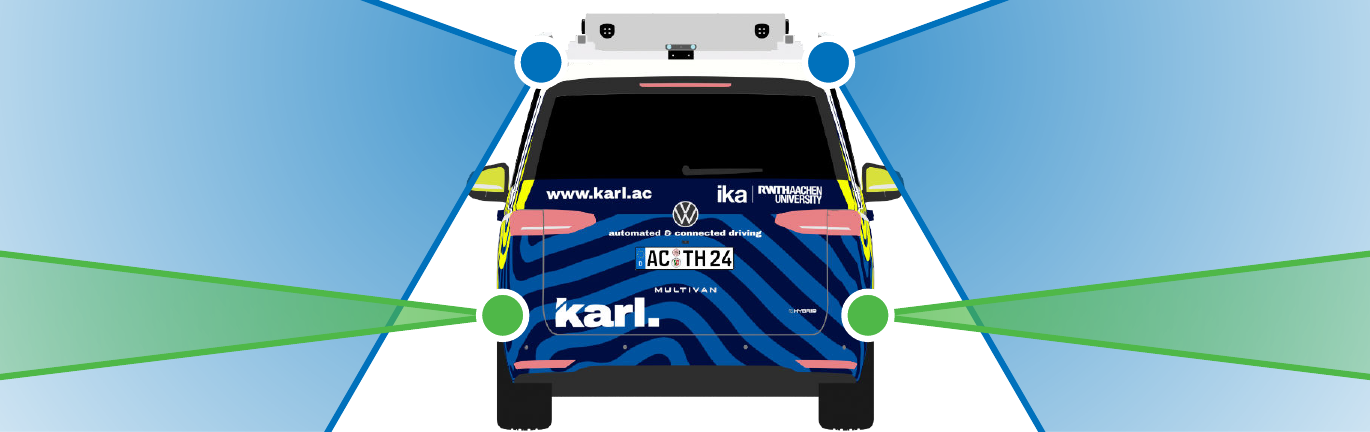}
    \caption{Rear view}
    \vspace{2mm}
    \includegraphics[width=\textwidth]{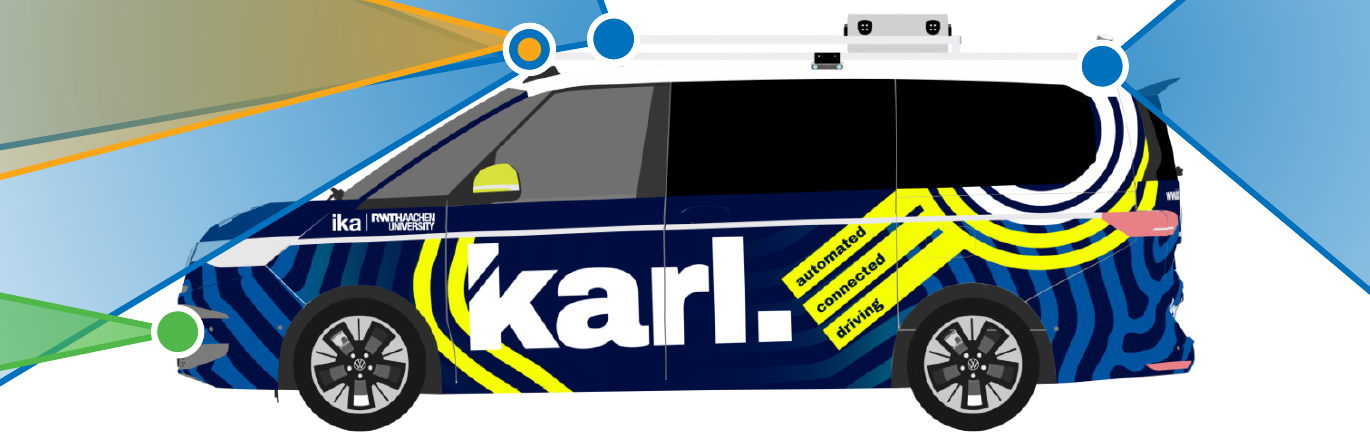}
    \caption{Side view}
    \vspace{2mm}
    \includegraphics[width=\textwidth]{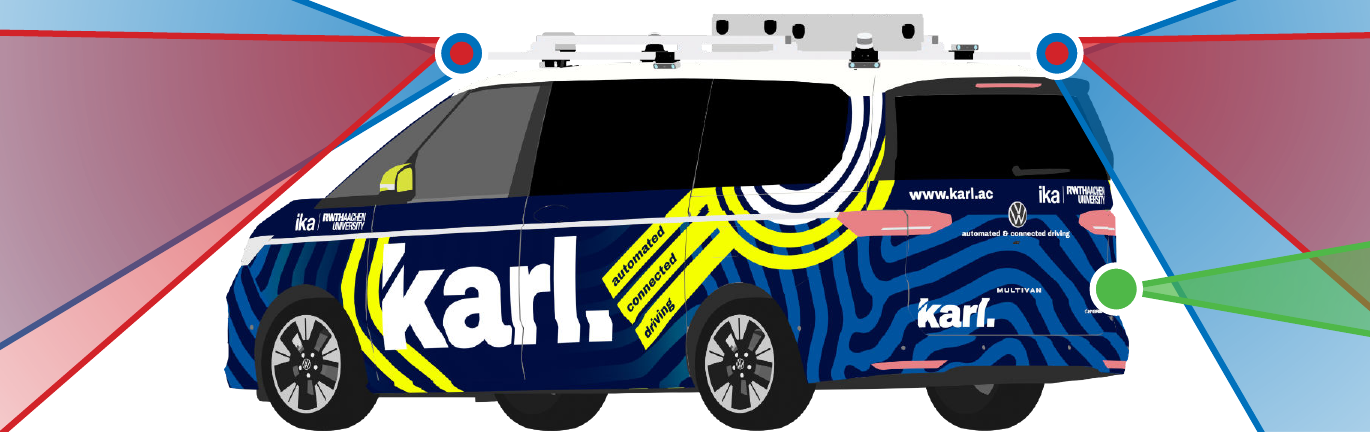}
    \caption{Rear three-quarter view}
  \end{subfigure}
  \caption{Fields of view of environment sensors: \protect\colorcircle{rwthblue} stereo cameras, \protect\colorcircle{rwthred} rotating lidars, \protect\colorcircle{rwthorange} FMCW lidar, and \protect\colorcircle{rwthgreen} 4D radars. The top-down view shows all sensors, the other views only show those sensors whose vertical field of view lies in the view plane.}
  \label{fig:sensor-fov}
\end{figure*}

Cameras and lidars are mounted on the sensor rack (\sect{sec:sensor-rack} and \fig{fig:sensor-rack}). The radars are integrated into the vehicle body, with four units located behind the front and rear bumpers, and one embedded in the front grill. A key criterion for sensor placement is to achieve \SI{360}{\degree} coverage with minimal blind spots, illustrated in \fig{fig:sensor-fov}. The essential characteristics of all primary environment sensors are summarized in~\tab{tab:sensors}.

Since our automated driving software stack (\sect{sec:ad-stack}) is based on \textit{ROS~2}~\cite{ros2}, the availability of official ROS drivers was a key criterion in sensor selection.

\begin{table}[b]
\centering
\caption{Characteristics of main environment sensors; italic script denotes configurable properties; resolution and FPS with max.\ data rate are given.}
\setlength{\tabcolsep}{5.0pt}
\begin{tabular}{@{}l
                S[table-format=3.0]
                S[table-format=2.0]
                r
                r
                S[table-format=2.0]@{}}
\toprule
\textbf{Sensor} & \textbf{Yaw [\textdegree]} & \textbf{Pitch [\textdegree]} & \textbf{FoV [\textdegree]} & \textbf{Resolution @ FPS [Hz]} \\ 
\midrule
\multicolumn{5}{@{}l}{\colorcircle{rwthblue} \textbf{Camera}} \\ 
front center      &      0 &   5  &  80×52  & \textit{1920x1200 @ 60} \\
front left/right  &  \pm45 &   5  &  80×52  & \textit{1920x1200 @ 60} \\
mid left/right    &  \pm90 &  20  & 110×80  & \textit{1920x1200 @ 60} \\
rear center       &    180 &   0  & 110×80  & \textit{1920x1200 @ 60} \\
rear left/right   & \pm135 &  20  & 110×80  & \textit{1920x1200 @ 60} \\
mid center        &      0 & -30  & 110×80  & \textit{1920x1200 @ 60} \\
\midrule
\multicolumn{5}{@{}l}{\colorcircle{rwthred}\colorcircle{rwthorange} \textbf{Lidar}} \\ 
front center     &      0 & 20 & \textit{120x29} & \textit{2000x64 @ 20} \\
front left/right &  \pm45 & 20 & 360×42 & \textit{1024x128 @ 20} \\
rear left/right  & \pm135 & 20 & 360×42 & \textit{1024×128 @ 20} \\
\midrule
\multicolumn{5}{@{}l}{\colorcircle{rwthgreen} \textbf{Radar}} \\ 
front center     &      0 & 0 & \textit{110x23} & \textit{- @ 15} \\
front left/right &  \pm90 & 0 & \textit{130×14} & \textit{- @ 15} \\
rear left/right  & \pm143 & 0 & \textit{100×20} & \textit{- @ 15} \\
\bottomrule
\end{tabular}
\label{tab:sensors}
\end{table}

\subsubsection{Cameras}
\label{sec:cameras}

To cover the full \SI{360}{\degree} environment, we choose nine roof-mounted \textit{StereoLabs ZED X} stereo cameras. The three front-facing cameras are equipped with \SI{4}{mm} focal length lenses, giving them a narrower horizontal field of view of \SI{80}{\degree} for enhanced resolution at larger distances. The five side- and rear-facing cameras instead rely on \SI{2}{mm} focal length lenses with \SI{110}{\degree} horizontal field of view. Another \SI{2}{mm} camera is pointed upwards for robust traffic light detection. All lenses use polarizing filters to reduce glare and reflections. The built-in \textit{ZED SDK} provides short-range 3D bounding boxes and depth estimations out-of-the-box.

All cameras are capable of running up to 1200p video at \SI{60}{fps}. They use GMSL2 connectors to reduce electromagnetic interference and enable high data rates. We dedicate two \textit{NVIDIA Jetson AGX Orins} (\sect{sec:embedded-ai-computers}) to solely process four camera streams each.

\subsubsection{Lidars}

The lidar setup consists of four rotating lidars and a front-facing solid-state lidar. Although the rotating ones individually cover the \SI{360}{\degree} environment, a single roof-mounted rotating sensor would be subject to large blind spots due to the vehicle dimensions. To minimize blind spots, we therefore place one \textit{Ouster OS1}, optimized for high-resolution point clouds at medium range, on each of the four sensor rack corners and tilt them downwards. The sensors can produce point clouds of up to \num{128} layers at a horizontal resolution of up to \num{4096} angular bins and a frequency of up to \SI{20}{Hz}, covering a \SI{360}{\degree} horizontal and a \SI{42}{\degree} vertical field of view.

The front-facing \textit{Aeva Aeries II} solid-state lidar is based on FMCW~(frequency-modulated continuous wave) technology. In addition to 3D position, reflectivity, and confidence, this sensor also returns the radial velocity for each measured point in space. With a \SI{10}{\percent} reflectivity range of up to \SI{200}{m} and a field of view of up to \SI{120}{\degree} horizontally and \SI{29}{\degree} vertically, the FMCW lidar allows us to look far into the region in front of the vehicle. The additional velocity measurement provides valuable information for perception and tracking tasks.

\subsubsection{Radars}
\label{sec:radars}

Radars provide weather‐robust and velocity-rich sensing, making them a key complement to our camera and lidar suite. On \textit{karl.}, we deploy three types of 4D imaging radars that, contrary to conventional radars, also resolve information vertically: one front-facing \textit{Altos V2} mounted below the license plate for long-range forward-facing perception; two side-facing \textit{smartmicro DRVEGRD169} units integrated under the left and right sections of the front bumper, providing wide-angle coverage of the vehicle's lateral sectors; and two rear-oriented \textit{smartmicro DRVEGRD152} units mounted under the rear bumper with a slight outward angle, covering both the rear and rear-to-the-side regions.

All radars output range, radial velocity, azimuth, elevation, and reflectivity. For example, the \textit{DRVEGRD169} supports multiple modes from ultra-short to long range of up to \SI{130}{m} and a field of view of up to \SI{130}{\degree} azimuth. The \textit{Altos V2} is reported to detect pedestrians at \SI{200}{m} and cars at \SI{400}{m} with an angular resolution below \SI{1.5}{\degree}.

\subsubsection{Sensor Calibration}

In order to relate 3D spatial information captured by the sensors to the vehicle, all environment sensors have to be properly calibrated extrinsically to the vehicle's frame of reference. We use the open-source calibration tool \texttt{ros2\_calib}~\cite{till_beemelmanns_2025_17218809} for the auto-calibration and manual finetuning process. The intrinsic calibration information of each camera lens is provided by the manufacturer.

\subsection{Localization}
\label{sec:localization}

To safely navigate environments ranging from open-sky highways to GNSS-impaired urban canyons, highly precise localization is required. As the main localization hardware unit, the common choice is to combine GNSS-based satellite positioning (Global Navigation Satellite System) with an INS~(Inertial Navigation System) composed of inertial sensors and with RTK~(Real-Time Kinematic Positioning) for real-time corrections, enabling centimeter-level accuracy.

We choose a compact and fully-integrated RTK/GNSS-INS system in the form of an \textit{SBG Systems Ekinox Micro}. The sensor promises a positioning accuracy of up to \SI{1}{cm} and a heading accuracy of up to \SI{0.05}{\degree}. Combined with two \textit{Tallysman VSP6037L} GNSS antennas on the sensor rack, the localization system is also able to measure the vehicle heading at standstill. The sensor itself is installed in the cabin rack, centered above the rear axle.

\subsection{External Communication \& V2X}
\label{sec:communication}

Reliable connectivity is essential for automated and connected vehicles. It not only provides RTK/GNSS correction data (\sect{sec:localization}), but also enables use cases such as remote operation, live digital twins, over-the-air updates, and function offloading. Besides cellular access, our platform supports standardized V2X (vehicle-to-everything) communication.

\subsubsection{5G Cellular Networking}

As the main gateway to the internet, we choose an \textit{Ericsson Cradlepoint R1900} automotive-grade 5G~router coupled with an \textit{RX30-MC} extension kit containing an \textit{MC400} modular modem. This dual-modem setup yields an integrated quad-SIM dual-active 5G router, i.e., two out of four SIM cards in total can be active at the same time, with the remaining two SIM cards acting as fallbacks. 

The router is installed in the cabin rack and connected to a \textit{Panorama Antennas LGMHM4B-6-60-24-58} 4x4-MIMO antenna mounted on top of the rooftop box.

In order to guarantee secure and bi-directional communication between vehicle and external infrastructure, e.g., cloud servers, we employ a private \textit{WireGuard} VPN.

\subsubsection{V2X Connectivity}
\label{ssec:v2x}

While the general-purpose 5G connectivity lets us research arbitrary types of communication with entities external of the vehicle, we install a dedicated \textit{Cohda Wireless MK6 OBU} for standardized V2X communication. The device is compatible with direct WiFi-based V2X (IEEE 802.11p / ETSI~ITS-G5 / DSRC) as well as direct and cellular C-V2X. Note that the base vehicle is natively publishing ETSI~ITS-G5 messages in addition to messages sent by our ADS. The V2X modem is installed in the cabin rack next to the 5G router.

\subsection{Compute \& Networking}
\label{sec:compute-networking}

Computation is at the heart of automated driving. Combined with the trend towards software-defined vehicles~(SDVs), E/E~architectures are shifting from distributed architectures to domain or zonal or even fully-centralized architectures~\cite{vehicle-architecture}. From an R\&D perspective, focusing on a single centralized high-performance computer handling all computational workloads has the additional benefit of less software orchestration overhead, leading to quick development cycles on the underlying software stack.

\subsubsection{High-Performance Computer (HPC)}
\label{ssec:hpc}

We install a custom-built, but otherwise regular high-performance workstation as the main centralized processing unit, running all parts of our automated driving software stack. The air-cooled computer is hosting an \textit{AMD Ryzen Threadripper PRO 9985WX} \texttt{x86-64} CPU, two \textit{NVIDIA RTX PRO 6000 Blackwell Max-Q} GPUs, and \SI{256}{GB} of DDR5-RAM. At the time of writing, it is running \textit{Linux Ubuntu~24.04}. From a research point of view, the main purpose of the powerful \num{64}-core~\num{3.2}/\SI{5.4}{GHz}~CPU, the high-end GPUs for deep learning applications, and the large memory is to avoid computation bottlenecks and keep headroom for future developments, e.g., towards end-to-end learning-based software stacks. Large data recording campaigns are enabled through \SI{8}{TB} of internal PCIe~5.0 NVMe M.2 SSD storage and multiple portable USB~3.2~Gen~2x2 SSDs.

\subsubsection{Embedded AI Computers}
\label{sec:embedded-ai-computers}

In addition to the HPC, we install two \textit{NVIDIA Jetson AGX Orin} developer kits in the rooftop box (\sect{sec:sensor-rack}). Together with a \textit{StereoLabs ZED Link Capture Card Quad}, they provide an interface to the GMSL2 cameras (\sect{sec:cameras}) that are optimized for being driven by \textit{NVIDIA Jetson Orin}. This way, we reduce the computational burden on the HPC by processing the heavy load of nine stereo videos (and optionally, built-in depth and object perception) directly on the embedded devices. 

Although the embedded AI computers currently only drive four cameras each, they also open the possibility for testing and validating software components of the ADS on embedded \texttt{arm64}-based hardware that is closer to a production-ready compute platform like \textit{NVIDIA DRIVE}.

\subsubsection{Networking \& Switches}
\label{sec:networking-switches}

\begin{figure}
    \centering
    \includegraphics[width=1.0\linewidth]{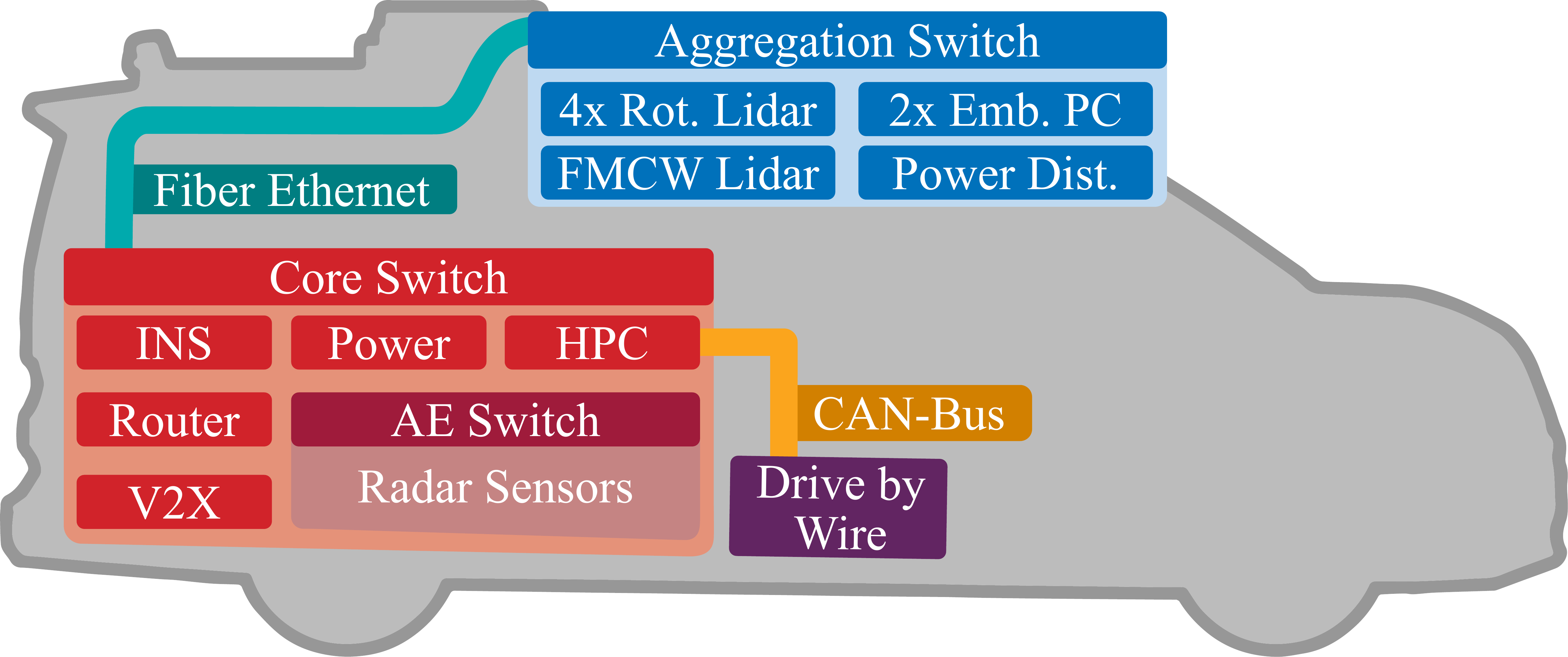}
    \caption{Overview of the two-tier Ethernet architecture connecting rooftop sensors and compute to the in-cabin core network}
    \label{fig:network}
\end{figure}

As shown in \fig{fig:network}, the system adopts a two-tier Ethernet architecture that separates rooftop sensor aggregation from in-cabin compute and communication devices.

A \textit{MikroTik CRS310-1G-5S-4S+IN} rooftop aggregation switch inside the rooftop box (\sect{sec:sensor-rack}) concentrates all five lidars, the two embedded computers, and the rooftop power distribution unit. The switch was selected for its DC supply and flexible SFP/SFP+ port layout, enabling direct integration of all rooftop lidars including the FMCW lidar via an Automotive-Ethernet-to-SFP converter.
The cabin rack hosts the core of the in-vehicle network.

A \textit{Cisco CBS350-48NGP-4X-EU} 48-port switch provides the high-bandwidth fiber link to the rooftop aggregation switch and distributes Ethernet connectivity to in-cabin devices. Its large port count and Power-over-Ethernet~(PoE+) budget leave substantial headroom for future compute or sensor expansion. Beyond linking the RTK/GNSS-INS unit (\sect{sec:localization}), the 5G router and V2X unit (\sect{ssec:v2x}), and the HPC (\sect{sec:compute-networking}), the core switch also interfaces with the power-distribution unit (\sect{ssec:power-supply}) for remote control of rooftop power channels.
A \textit{Technica Engineering Enhanced Ethernet Switch MateNET} attaches to the core switch to aggregate Automotive-Ethernet radar streams before forwarding them to the HPC. All switches offer full management capabilities and allow fine-grained control over link settings, VLANs, and bonding for diverse experiment and data collection setups. For development convenience, the core network additionally exposes selected ports on the cabin rack top and in the central vehicle console.

\subsubsection{PTP Clock Synchronization}
\label{ssec:ptp}

Accurate timestamp alignment across all sensors and compute units is essential for data fusion, replay, and time-critical tasks. \textit{karl.} therefore uses IEEE~1588 Precision Time Protocol (PTP) throughout the in-vehicle network. The GNSS-disciplined INS (\mbox{\sect{sec:localization}}) acts as the PTP grandmaster, providing a stable satellite-synchronized UTC reference clock.

PTP is distributed over the two-tier Ethernet topology (\sect{sec:networking-switches}). As the installed switches do not support transparent clocking, we operate PTP in end-to-end mode, which still provides sufficient accuracy for our sensor suite and compute stack. The HPC and both embedded computers synchronize their NIC hardware clocks via \texttt{linuxptp}. Sensors with native PTP support, such as the \textit{Ouster} lidars and the \textit{Cohda Wireless} V2X unit, lock directly to the same PTP domain. Devices without PTP capability inherit synchronized time from their host: the camera images are timestamped on the \textit{Orins} and the radar data streams are timestamped on the HPC. While the \textit{Aeva Aeries II} lidar supports PTP, it is not compatible with our chosen PTP configuration. Consequently, we provide it with synchronized time via an NTP service distributed by the HPC.

\subsection{Vehicle Interface \& Drive-by-Wire System}
\label{sec:dbw-system}

Taking over a series production vehicle's longitudinal and lateral control is arguably the most important modification for automating the driving task. To this end, installation of external actuators (e.g., bulky steering robots) or an intrusion of existing vehicle actuators (via vehicle-specific bus systems and protocols) is possible. Stock vehicle sensors may also be interfaced, but generally have to be augmented through additional sensors anyway.

We have chosen a drive-by-wire system~(DbW) by \textit{fka~GmbH}\footnote{\textit{\href{https://www.fka.de/en/competences/automated-driving/highlights/196-our-vehicle-fleet-for-tests-on-public-roads.html}{https://www.fka.de/en/competences/automated-driving}}}, implemented using a \textit{dSPACE MicroAutoBox III} and a \textit{PEAK PCAN-Router FD}, acting as interfaces to the vehicle CAN bus. An additional CAN bus is used as an interface to the HPC, on which selected signals can be read and requests for vehicle control can be sent.
For lateral vehicle control, the system provides a curvature-based interface using the vehicle's active parking assist.
For longitudinal control, the system interfaces the vehicle's ACC, providing a target acceleration interface.
Finally, interfaces for secondary controls, such as the turn signals, are available. The system can be activated using the steering wheel buttons and an additional operator HMI (\sect{ssec:dev-hmi}). The DbW system provides safety mechanisms to ensure controllability of the vehicle by a safety driver. These include immediate driver override capabilities, an emergency stop switch, steering-rate limitations, and key-based access restrictions.

\section{ADS Software Stack}
\label{sec:ad-stack}

The vehicle setup presented in the previous section is laying the foundation for making \textit{karl.}\ a flexible and powerful vehicle platform for automated and connected driving research. The software that runs on such a vehicle is at least of equal importance as the hardware. Although more details are out-of-scope here, this section provides some insights into the software stack of our automated driving system~(ADS) and the development process surrounding it.

\subsection{DevOps \& Containerization}
\label{sec:devops-and-containerization}

To enable quick development cycles, we implement \mbox{DevOps} best practices along the continuous software life cycle. These span from infrastructure-as-code practices~(Ansible), over version-controlling source code of all ADS software components~(Git), automatically building and testing the source code in CI/CD pipelines, multi-layer testing in simulation, deploying modules as reproducible container images, mindful monitoring and recording of system performance, all the way to closing the DevOps cycle and iterating on the source code again.

Testing and validating ADS software components on the vehicle constitutes system-level end-to-end testing rather than unit or integration testing. In addition to functional validation, the vehicle also supports demonstration activities and high-quality data collection. For these reasons, we decouple source code development from stack deployment in the vehicle. Instead of cloning and building the ROS~2-based source code, we use \texttt{docker-ros}~\cite{docker-ros} to automatically build ready-to-deploy container images in CI/CD.

We refer to the collection of containers and their configurations as \textit{compositions}, which we then operate on the vehicle using \textit{Docker Compose}.

\subsection{Operation Modes}

The modes of operation that \textit{karl.}\ supports depend entirely on the combination of activated components in the software composition and activated hardware interfaces. From manually driving the vehicle in its unmodified series-production state~-- when the added power supply is off and the drive-by-wire actuator interface is physically disconnected from the vehicle~-- to L4 self-driving powered by our automated driving stack, it is up to the vehicle operators to flexibly adjust the system under test.

We primarily distinguish between the following operation modes that we can switch between dynamically:
\begin{enumerate}
    \item \textbf{Stock:} All additional hardware is disconnected from the power supply; DbW is physically separated from CAN; vehicle is driven by human driver; used for vehicle transport.
    \item \textbf{Shadow:} Selected sensors and software components are active; DbW is physically separated from CAN; vehicle is driven by human driver; used for data recording and open-loop shadow mode testing.
    \item \textbf{Automated / Remote Control:} Selected sensors and software components are active; DbW system is engaged and exerting lateral and longitudinal control; human safety driver can manually override at any time; used for testing, validating, and demonstrating the automated driving and remote operation capabilities on test tracks and soon in public road traffic.
\end{enumerate}

\subsection{Automated Driving Stack}
\label{sec:ads}

\begin{figure}[b]
    \centering
    \includegraphics[width=1.0\linewidth]{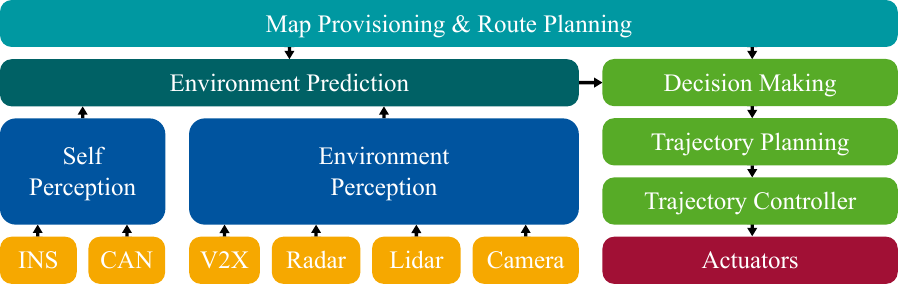}
    \caption{Fundamental components of our software stack for automated and connected driving}
    \label{fig:stack}
\end{figure}

In order to operate \textit{karl.} in L4 automated mode, we use our self-developed software stack for automated and connected driving~(AD stack). As described in \sect{sec:devops-and-containerization}, it is fully implemented in ROS~2 and containerized at software component level. We only briefly describe the stack's components here and in~\fig{fig:stack} to support the notion of \textit{karl.'s} capabilities.

At the lowest layer, ROS~2 drivers provide standardized interfaces to all sensor and actuator components. We integrate standardized ETSI~ITS-G5 V2X messages received from, e.g., many \textit{Volkswagen Group} series production vehicles~\cite{v2aix}, to complement the environment information gathered with the vehicle's sensors.

On top of this layer, perception components process data from all sensors to create an accurate and consistent representation of the environment. This includes, e.g., object detection in lidar point clouds, traffic light detection in camera images, as well as self-perception using our INS and the vehicle's CAN.

The subsequent prediction and behavior planning modules use the processed environment information and high-definition map data to predict the behavior of other traffic participants and derive the vehicle's own behavior decision.

Based on these decisions, an optimal control problem (OCP) is solved to generate a feasible trajectory that respects the vehicle's dynamic capabilities and adheres to constraints derived from traffic rules and other road participants. For this purpose, vehicle motion is modeled according to the single-track model with Ackermann steering.
A subsequent controller then provides acceleration and curvature to the actuator interface.

We plan to publish further details about the AD stack, including architecture, interfaces, and individual module designs, in a dedicated publication.

\section{Evaluation}

We have developed \textit{karl.}\ to be a comprehensive, flexible, and powerful research platform for many kinds of research, engineering, and validation in the context of automated and connected driving. In this section, we are evaluating some core capabilities of the vehicle. Note that we focus on the vehicle setup\footnote{The evaluation was conducted with an earlier HPC setup including an AMD Ryzen Threadripper PRO 5995WX and one NVIDIA GeForce RTX 4090.}, an evaluation of the software stack is out of scope.

\subsubsection{Clock Synchronization}

For an evaluation of proper clock synchronization between HPC and embedded computers, we measure the PTP~clock offset to the GNSS~PTP~grandmaster \num{30} times over the course of \SI{3}{min} using \texttt{pmc}. For all devices, the mean offset is below \SI{200}{ns}. The same sub-microsecond accuracy is expected to hold for the PTP-synchronized sensors.

\subsubsection{Sensor Coverage \& Calibration}

The four tilted corner lidars minimize blind spots in point cloud-based sensing. The FMCW lidar covers a long range in front of the vehicle, including velocity estimation. Similarly, the radars cover relevant areas with a lower mounting point. The cameras further minimize the lidar blind spots and are suited for close-range perception. As depicted in~\fig{fig:sensor-fov}, all main sensor modalities cover the \SI{360}{\degree} environment in a redundant manner.

\begin{figure}[t]
  \centering
  \begin{subfigure}[b]{\linewidth}
    \includegraphics[width=\linewidth]{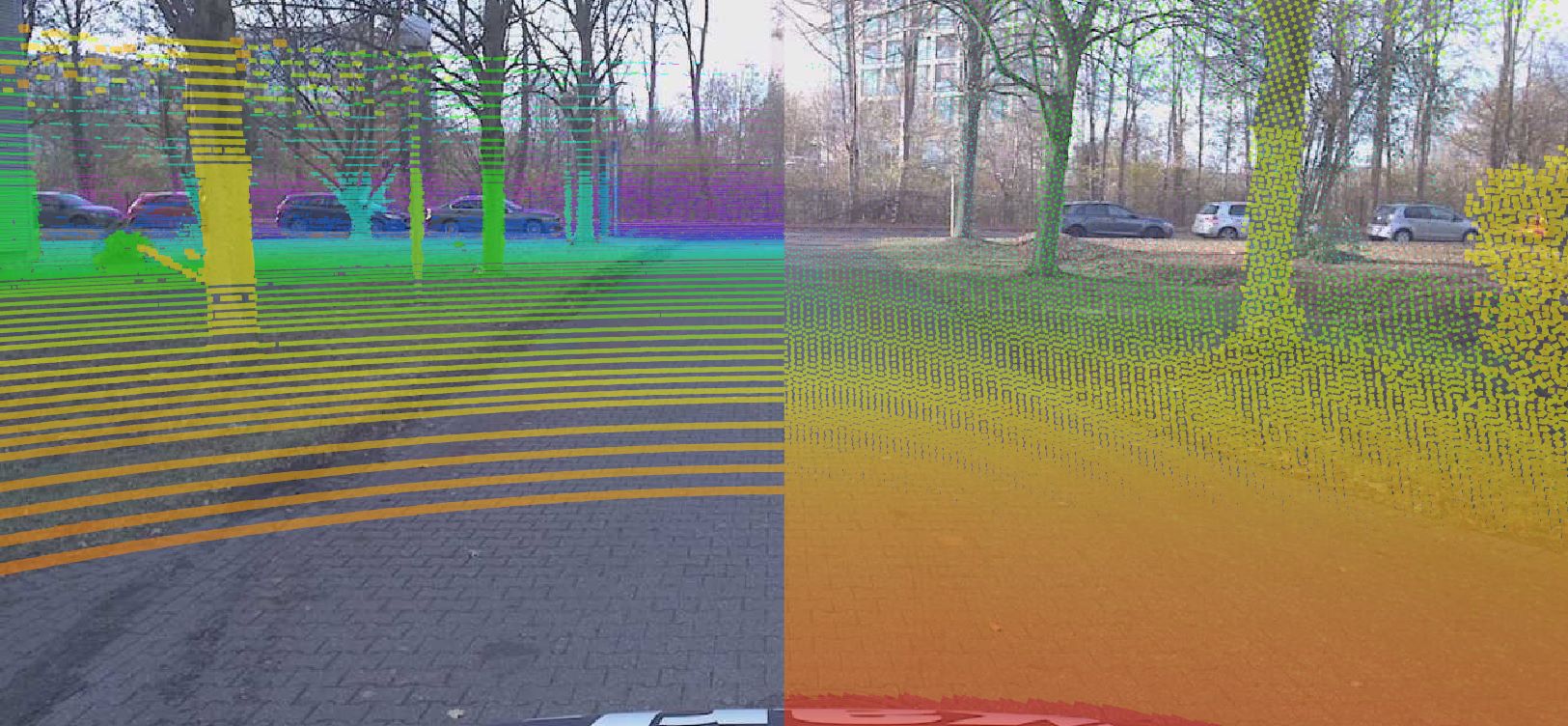}
    \caption{FMCW lidar's (left) and front rotating lidars' point clouds projected onto front center camera image}
    \medskip
  \end{subfigure}
  \begin{subfigure}[b]{\linewidth}
    \includegraphics[width=\linewidth]{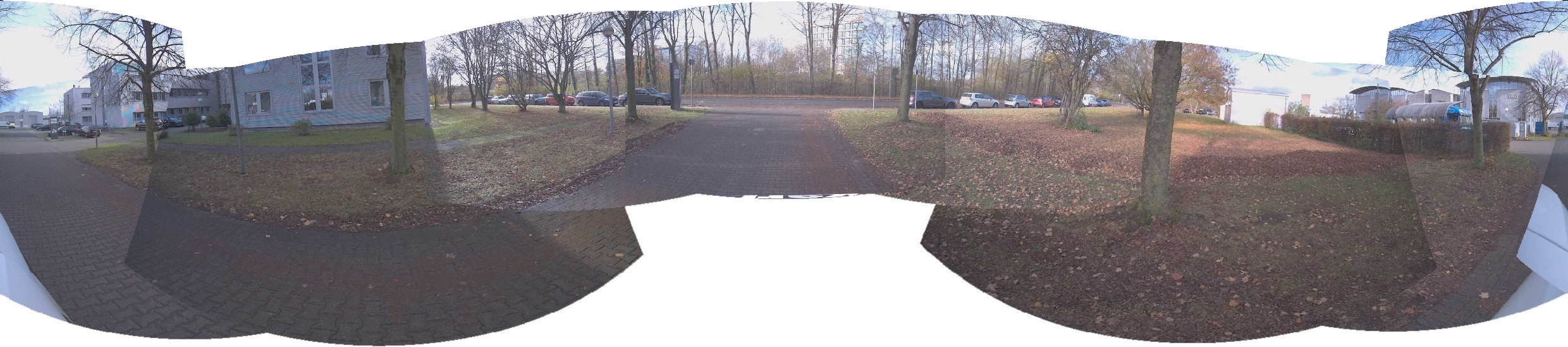}
    \caption{Equirectangular \SI{360}{\degree} projection using eight cameras}
  \end{subfigure}
  \caption{Sample demonstrating sensor calibration quality at rest}
  \label{fig:sensor-calibration}
\end{figure}

We qualitatively evaluate the sensor calibration between cameras and lidars at rest. \fig{fig:sensor-calibration} shows good projection accuracy in the near and far field. Note that the equirectangular camera reprojection can mathematically only match perfectly at one particular distance.

\subsubsection{Sensor Data Latency}

ROS~2 drivers form the entry point to the AD stack. We measure the mean latency from sensor data capture to ROS~2 message availability in the AD stack. We compress images and point clouds directly in the driver. 

For the rotating lidars, the PTP-synced timestamp is set at the start of each \SI{360}{\degree} rotation. At \num{1024}$\times$\num{128}@\SI{20}{Hz}, the mean latency to ROS message availability is \SI{72}{ms}, comprising \SI{50}{ms} for the sensor sweep, \SI{20}{ms} for internal driver processing, and \SI{2}{ms} for ROS communication.

For the FMCW~lidar at \num{2000}$\times$\num{64}@\SI{20}{Hz}, mean latency from NTP-synced timestamp to ROS is \SI{220}{ms}, with \SI{219}{ms} from internal processing and \SI{1}{ms} from ROS.

For the stereo cameras at \num{720}p@\SI{15}{Hz}, we measure \SI{139}{ms} mean latency, including \SI{120}{ms} driver and \SI{19}{ms} ROS latency. Radar sensors show negligible latencies below \SI{1}{ms} due to minimal data volume.

\subsubsection{Power Budget}

When all hardware components (HPC, sensors, etc.) are turned on, but idle, the total power consumption of the system amounts to \SI{\sim 660}{W}. Under full load, i.e., during operation of the full AD stack, the power consumption is \SI{\sim 1100}{W}.

Based on the available battery capacity (\sect{ssec:power-supply}), the system can therefore be operated under full load for approximately \SI{\sim 4.5}{h} without additional battery charging. However, since the system can be charged with up to \SI{90}{A} from the vehicle battery, the operating time of the system, in practice, is only limited by the PHEV's internal energy storage.

\subsubsection{Compute Headroom}

Under full load of our current AD stack, we still have headroom for additional computational demands. On the HPC, over \SI{2}{min}, we measure an average system load of \SI{80}{\%}~CPU, \SI{30}{\%}~GPU, and \SI{8}{\%}~RAM. On the embedded computers, we measure \SI{37}{\%}~CPU, \SI{24}{\%}~GPU, and \SI{18}{\%}~combined memory.

\subsubsection{Drive-by-Wire System}

We have conducted several experiments to test the limits of the vehicle control interface.
We provide nominal values for acceleration and curvature from our ROS interface to the DbW system, receive measured values back in ROS, and compare nominal vs.\ real values.
We measure longitudinal acceleration with the INS at \SI{50}{Hz} and compute a centered rolling average over \SI{200}{ms}, and measure steering angle via the curvature interface of the DbW system at \SI{50}{Hz}.

\begin{figure}[t]
  \centering
  \includegraphics[width=1.0\linewidth]{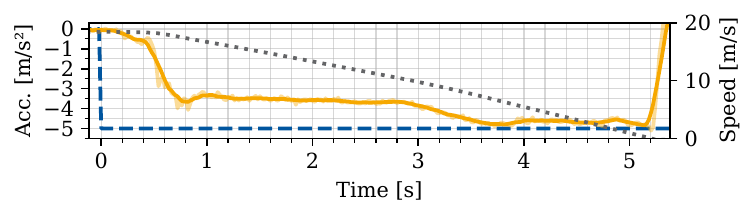}
  \caption{\protect\colorcircle{rwthorange} Vehicle response to \protect\colorcircle{rwthblue} nominal braking step with \protect\colorcircle{rwthgrey} vehicle speed}
  \label{fig:acc-step}
\end{figure}

Since the DbW system is based on the stock ACC longitudinally, we expect it to act within the limits dictated by ISO~15622~\cite{ISO15622}. We can confirm that the DbW system reaches and adheres to the longitudinal acceleration limits of \SI{-3.5}{m/s^2} and \SI{-5.0}{m/s^2} (below \SI{5}{m/s}) as illustrated in \fig{fig:acc-step}.

Laterally, based on a safety assessment, we apply a speed-dependent limitation on maximum curvature and curvature rate of the stock parking assist. Again, we can confirm that the DbW system reaches and adheres to these limits.

\subsubsection{Regulatory Approval}

The final step towards making \textit{karl.}\ a comprehensive research platform for automated and connected driving is to bring it to public roads. We are currently in the process of obtaining an official testing approval for German public road traffic in accordance with §1i~StVG in conjunction with §16~AFGBV.
We comply with data protection obligations via a dedicated privacy concept and clear informational signage on the vehicle.

\section{Conclusion}

We have presented \textit{karl.}, our new research vehicle for automated and connected driving. We have positioned \textit{karl.}\ as a capable, comprehensive, extensible, and future-proof research platform for many kinds of research in the field: from L4 ADS development and validation over remote operations and infrastructure support to user experience and acceptance studies. The technical details presented aim to share knowledge about building such a vehicle platform for research by providing insights into all relevant design decisions. Our evaluation demonstrates that \textit{karl.}\ fulfills many required core capabilities. Looking forward, we are currently focusing on obtaining a testing approval for public road traffic and thereby laying the foundation to operate \textit{karl.}\ for many years to come.

\section*{ACKNOWLEDGMENTS}

Building and actively maintaining a research vehicle is hard: running our own automated driving stack on \textit{karl.}\ is only made possible through the work of many individuals in our \textit{Vehicle Intelligence \& Automated Driving} department. We thank all past, current, and future involved colleagues.

This work was funded by the German Federal Ministry of Research, Technology and Space (BMFTR) in projects \textit{6GEM+} (16KIS2409K), \textit{6GEM} (16KISK036K), \mbox{\textit{autotech.agil}} (1IS22088A); by the European Union under the Horizon Europe programme in projects \textit{AIGGREGATE} (101202457), \textit{AIthena} (101076754), \textit{SYNERGIES} (101146542); and by the German Research Foundation~(DFG) in project \textit{4-CAD} (503852364).

\bibliographystyle{IEEEtran}
\bibliography{root} 
	
\end{document}